\documentclass[%
 reprint,
 superscriptaddress,
 amsmath,amssymb,
 aps,
 prl
]{revtex4-2}

\usepackage{graphicx} 
\usepackage{hyperref} 
\usepackage{dcolumn} 
\usepackage{bm} 
\usepackage[dvipsnames]{xcolor} 
\usepackage{braket} 
\usepackage{cancel} 
\usepackage{soul} 
\usepackage[capitalize]{cleveref} 
\usepackage[all]{hypcap} 

\DeclareMathOperator{\arsinh}{arsinh}

\begin{document}

\title{
Mechanical cosmology: simulating scalar fluctuations in expanding Universes using synthetic mechanical lattices
}

\author{Brendan Rhyno}
\email{brhyno2@illinois.edu}
\affiliation{Department of Physics, University of Illinois at Urbana-Champaign, Urbana, Illinois 61801, USA}

\author{Ivan Velkovsky}
\affiliation{Department of Physics, University of Illinois at Urbana-Champaign, Urbana, Illinois 61801, USA}

\author{Peter Adshead}
\affiliation{Illinois Center for Advanced Studies of the Universe \& Department of Physics, University of Illinois at Urbana-Champaign, Urbana, Illinois 61801, USA}

\author{Bryce Gadway}
\affiliation{Department of Physics, University of Illinois at Urbana-Champaign, Urbana, Illinois 61801, USA}
\affiliation{Department of Physics, The Pennsylvania State University, University Park, Pennsylvania 16802, USA}

\author{Smitha Vishveshwara}
\email{smivish@illinois.edu}
\affiliation{Department of Physics, University of Illinois at Urbana-Champaign, Urbana, Illinois 61801, USA}

\date{\today}

\begin{abstract}
Inspired by recent advances in observational astrophysics and continued explorations in the field of analog gravity, we discuss the prospect of simulating models of cosmology within the context of synthetic mechanical lattice experiments. We focus on the physics of expanding Universe scenarios described by the Friedmann-Lema{\^i}tre-Robertson-Walker (FLRW) metric. Specifically, quantizing scalar fluctuations in a background FLRW spacetime leads to a quadratic bosonic Hamiltonian with temporally varying pair production terms.
Here we present a mapping that provides a one-to-one correspondence between these classes of cosmology models and feedback-coupled mechanical oscillators.
As proof-of-principle, we then perform experiments on a synthetic mechanical lattice composed of such oscillators.
We simulate two different FLRW expansion scenarios with Universes dominated by vacuum energy and matter and discuss our experimental results.
\end{abstract}

\maketitle

As current day probes of the cosmos, such as the James Webb Space Telescope \cite{Gardner:2006ky}, begin to reveal extraordinarily deep glimpses of the infant Universe, they lead to new questions on our present understanding of its beginnings. Inflation \cite{Guth:1980zm, Starobinsky:1980te,  Linde:1981mu,Albrecht:1982wi, Linde:1983gd} is now well established as the leading mechanism for setting the initial state of our Universe. An early phase of accelerated expansion, inflation was proposed to solve the horizon and flatness problems of the hot Big Bang cosmology. However, it was soon realized that quantum mechanical fluctuations of the  metric and fields during this epoch would generate nearly scale-invariant density and gravitational wave spectra \cite{Mukhanov:1981xt, Guth:1982ec, Hawking:1982cz, Bardeen:1983qw}.
Hand-in-hand with theoretical models, computational techniques, and observational astronomy, a surge in the development of analog gravitational systems serves to test, corroborate, and enhance this understanding \cite{Visser2002,Barcelo2011,Jacquet2020}.
Here, we show that synthetic mechanical lattices composed of measurement-based feedback-coupled mechanical oscillators are excellently poised to simulate key features of inflationary cosmology and, more generally, Friedmann-Lema{\^i}tre-Robertson-Walker (FLRW) cosmology.

The essence of the inflationary paradigm that lends itself to these experimental simulations is as follows.
Quantum fluctuations about a Bose-condensed inflaton field lead to density perturbations in the post inflationary Universe that clump and collapse under the influence of local gravity to give rise to the anisotropies in the cosmic microwave background, and the large scale structures in the Universe today. These quantum mechanical fluctuations arise as the zero-point motion of the fields which are then stretched to super-horizon scales due
to accelerated expansion. This quantum mechanical origin of structure, reflected in dynamical boson pair production, has become a crucial prediction of inflationary cosmology.

\begin{figure}[htp!]
    \includegraphics[width=0.45\textwidth]{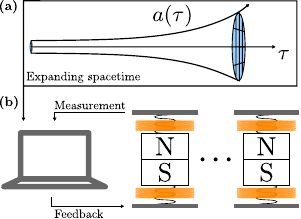}
    \caption{
    (Color online)
    Analog cosmology using coupled mechanical oscillators.
    (a) A cartoon depiction of the scale factor that dictates the expansion of a model Universe.
    (b) A synthetic mechanical lattice consisting of modular mechanical oscillators.
    Each oscillator consists of two springs holding a mass that is equipped with an accelerometer and a dipole magnet embedded in a pair of anti-Helmholtz coils.
    Input from the functional form of the scale factor,
    along with real-time accelerometer measurements,
    allows one to engineer feedback forces using the anti-Helmholtz coils which effectively makes the system behave as an analog simulator of the equations of motion in the cosmological theory.
    Analog versions of the correlation functions of interest in cosmology can then be extracted from the mechanical oscillator experimental data.
    }
    \label{fig:mechanical_cosmology}
\end{figure}

Historically, classical and quantum fluids have provided fertile ground for analog gravitational models ranging from black holes to the expanding Universe \cite{Unruh1981,Visser1998,Novello2002,Volovik2009,Barcelo2011}.
For instance, in Bose-Einstein condensates, in the hydrodynamic approximation, phase fluctuations obey the massless Klein-Gordon equation subject to an effective metric \cite{Garay2000,Garay2001,Jain2007}.
The versatile setting of ultra-cold atomic gases provides close cosmological parallels \cite{Fedichev2004_1,Fischer2004_1,Fischer2004_2,Hung2013,Schmiedmayer2013,Feng2018,Hu2019,Chatrchyan2021,Wilson2022,Steinhauer2022,TolosaSimeon2022,Viermann2022} even including protocols involving physical expansions \cite{Fedichev2003,Zoest2010,Prain2010,Cha2017,Eckel2018,GomezLlorente2019,Eckel2021,Sheehy2021,Banik2022,Carollo2022,Gaaloul2022,Bhardwaj2024}.
A plethora of alternatives exist for cosmological analogs, including in the realms of earth sciences \cite{Faraoni2022}, conducting wires \cite{Choudhury2019}, as well as metamaterials \cite{Leonhardt2006,Smolyaninov2012,Figueiredo2017} where customizable Hamiltonian dynamics can be engineered \cite{Riechert2022,Brandenbourger2019}.

Mechanical systems have evolved to an extraordinary level of sophistication in the gravitational context, finding applications in gravitational wave detectors and searches for dark matter and dark energy candidates \cite{Abbott2016,Vermeulen2021,Arvanitaki2016,Singh2020,Carney2021,Singh2021,Singh2022}. Here, we propose that they even provide a superb arena for analog gravity.
Our main observation is that in FLRW theories, the equations of motion of the bosonic degrees of freedom are ultimately those of coupled harmonic oscillators.
The unique feature of the oscillator network considered here is its access to highly tunable feedback forces \cite{Anandwade2023,Singhal2023,Tian2024}. 
When precise feedback forces can be engineered, we show there exists a natural mapping in which classical
oscillator systems are well positioned to perform analog simulations that directly target key properties of FLRW cosmology.
Here, we provide a proof-of-principle experimental demonstration in which we perform a comprehensive study of FLRW models in a synthetic mechanical lattice, simulating expansion scenarios driven by dark energy or matter, and obtaining the analogs of astrophysically relevant quantities, such as the power spectrum and the pair production.

In what follows we outline standard features of FLRW cosmology and then introduce the mechanical oscillator experiment pinpointing the parallels between the two. Finally, we present and discuss our experimental analog simulation results.

\textit{Conceptual aspects of FLRW cosmology.} We consider the simplest effective model that captures salient features of the quantum mechanical production of fluctuations in FLRW cosmology; namely, a free real massless scalar field $\varphi$ in a background spacetime ($\hbar=c=1$):
\begin{align}
    \label{eq:S}
    S &= \int d^4 x \sqrt{-g} \, \left( \frac{1}{2} g^{\mu\nu} \partial_\mu \varphi \partial_\nu \varphi \right),
\end{align}
where $g_{\mu\nu}$ denotes the components of the metric tensor with determinant $g$ and inverse $g^{\mu\nu}$. Here, the field $\varphi$ represents the fluctuations about the background inflaton field. (Alternatively, $\varphi$ could represent fluctuations in the transverse-traceless part of the spatial metric---gravitational waves.) 
Appropriate for large length scales in which the Universe is spatially homogeneous and isotropic, we focus on the (spatially-flat) FLRW metric:
\begin{align}
{\rm d}s^2 = a^2(\tau)({\rm d}\tau^2 -  {\rm d}\bm{x}^2),
\end{align}
where $a$ is the scale factor, and $\tau$ is conformal time. As is convention, primes represent derivatives with respect to conformal time, while overdots represent derivatives with respect to cosmic time. The two are related by $dt = ad\tau$.

The evolution of the scale factor follows from the Friedmann equations \cite{Friedmann1922}
\begin{align}
3M_{\rm Pl}^2 H^2 = \rho, \quad 
M_{\rm Pl}^2 \dot{H}  = -\frac{1}{2}\rho(1+w),
\end{align}
where $H = \dot{a}/a$ is the Hubble parameter, $M_{\rm Pl}$ is the Planck mass, and $\rho$ and $w = p/\rho$ are the energy density and equation of state of the matter fields driving the expansion of the Universe. For vacuum energy-dominated expansions, relevant for early Universe inflationary cosmology as well as the current state of our own Universe, the Hubble parameter is constant. In this case, the scale factor grows exponentially, $a\sim e^{H t}$.
For either a matter-dominated ($w = 0$) or radiation-dominated ($w = 1/3$) expansion, the scale factor exhibits power-law scaling with time: $a\sim t^{2 / (3(1+w))}$.

For our purposes, the quantum mechanical production of fluctuations in the scalar field in \cref{eq:S} about these expanding backgrounds is most easily analyzed by transforming from the Lagrangian to the Hamiltonian and canonically quantizing the theory.
Particle production can then be seen from the evolution of the creation and annihilation operators describing the instantaneous occupation of the initial vacuum state.
To quantize the theory, one can canonically normalize the kinetic term in \cref{eq:S}, by rescaling the physical field to $y(\tau,\bm{x})\equiv a(\tau) \varphi(\tau,\bm{x})$ \cite{Polarski1996,Burgess2023}. One then moves to a Hamiltonian description of the dynamics by introducing the canonical momentum, and performing a Legendre transformation. Canonical quantization amounts to promoting the comoving field and its canonical momentum to quantum field operators obeying canonical commutation relations.
By virtue of translational invariance, one Fourier transforms and introduces bosonic creation (annihilation) operators, $\hat b_{\bm{k}}^\dag$ ($\hat b_{\bm{k}}$),
in the standard way to reach the final form of the Hamiltonian operator \cite{Polarski1996}:
\begin{align}
    \label{eq:H}
    \hat{H}(\tau)
    =
    \int \frac{d^3 k}{2}
    &\bigg[
    k
    \left(\hat b^\dag_{\bm{k}} \hat b_{\bm{k}} + \hat b_{-\bm{k}} \hat b^\dag_{-\bm{k}}\right)
    \nonumber\\
    &+
    i\frac{a'(\tau)}{a(\tau)}
    \left( \hat b^\dag_{\bm{k}} \hat b^\dag_{-\bm{k}} - \hat b_{-\bm{k}} \hat b_{\bm{k}} \right)
    \bigg]
    ,
\end{align}
where $k=|\bm{k}|$ and $a'/a\ge0$ for the expansion scenarios considered here.
For a time-varying scale factor ($a'\neq0$), the instantaneous spectrum of the Hamiltonian is time-dependent and becomes unbounded from below for modes with $k<a'/a$
\cite{Nagel1995,Anglin2003,Rossignoli2005,Supplemental_Material}.

The Heisenberg equations of motion for the bosonic modes that follow from this Hamiltonian can be solved in terms of a Bogoliubov transformation \cite{Bogoliubov1947}:
$\hat b_{\bm{k}}(\tau) = u_{k}(\tau) \hat b_{\bm{k}} + v^*_{k}(\tau) \hat b^\dag_{-\bm{k}}$, where the Bogoliubov coefficients $u_{k}(\tau)$ and $v_{k}(\tau)$ evolve in conformal time according to the Bogoliubov equations:
\begin{align}
    \label{eq:Bogo_EOM}
    i
    \begin{pmatrix}
    u_{k}(\tau)\\
    v_{k}(\tau)
    \end{pmatrix} '
    &=
    \begin{pmatrix}
    k & i \frac{a'(\tau)}{a(\tau)} \\
    i \frac{a'(\tau)}{a(\tau)} & -k
    \end{pmatrix}
    \begin{pmatrix}
    u_{k}(\tau)\\
    v_{k}(\tau)
    \end{pmatrix}
    .
\end{align}
The $2\times 2$ matrix generating the dynamics is non-Hermitian and has imaginary instantaneous eigenvalues whenever $k<a'/a$ and real eigenvalues otherwise.
Matrices with this structure and unitary transformations of them appear frequently in the study of non-Hermitian quantum systems \cite{Bender2002,Bender2005,Guo2009,Xiao2019,Okuma2020,Gao2024}.
Note that these Bogoliubov equations are simply parametric oscillators in disguise \cite{Martin2019}. If one defines the function $Y_{k}(\tau) \equiv [ u_{k}(\tau) + v_{k}(\tau) ] / \sqrt{2k}$, it is straightforward to see that $0 = Y_{k}'' + ( k^2 - a''/a ) Y_{k}$. In the cosmological context, this function appears as the coefficients in the Fourier modes of the comoving field operator, $\hat y_{\bm{k}}(\tau) = Y_{k}(\tau) \hat b_{\bm{k}} + Y^*_{k}(\tau) \hat b^\dag_{-\bm{k}}$, and is thus intimately related to the power spectrum of the field.

The Bogoliubov equations and their solutions offer a convenient formalism to analyze important physical features associated with each expansion scenario.
For instance, under unitary time evolution generated by the Hamiltonian, \cref{eq:H}, the vacuum state $\ket{0}$ evolves into a two-mode squeezed state for each $(\bm{k},-\bm{k})$ pair \cite{Albrecht1994,Supplemental_Material}:
\begin{align}
    \label{eq:squeeze}
    \ket{\psi(\tau)} \propto \exp\left(
    \int \frac{d^3 k}{2} \, \tanh(r_{k}(\tau)) e^{i \theta_{k}(\tau)} \hat b^\dag_{\bm{k}} \hat b^\dag_{-\bm{k}}
    \right) \ket{0}
    ,
\end{align}
with squeeze parameter $r_{k}(\tau) \equiv \arsinh(|v_{k}(\tau)|)$ and squeeze angle
$\theta_{k}(\tau) \equiv \arg(u_{k}(\tau))-\arg(v_{k}(\tau))$.

It is also straightforward to calculate vacuum expectation values of Heisenberg operators in terms of the Bogoliubov coefficients.
At conformal time $\tau$, the vacuum expectation value for the number of excitations in a given $\bm{k}$ mode, the number of particle pairs produced with opposite momenta, and the power spectrum of the comoving field are determined, respectively, by:
\begin{subequations}
\label{eq:2pt}
\begin{align}
    \label{eq:particle_number}
    \braket{0| \, \hat b^\dag_{\bm{k}}(\tau) \, \hat b_{\bm{k}}(\tau) \, |0}
    &=
    |v_{k}(\tau)|^2
    ,\\
    \braket{0| \, \hat b^\dag_{\bm{k}}(\tau) \, \hat b^\dag_{-\bm{k}}(\tau) \, |0}
    &=
    u^*_{k}(\tau) \, v_{k}(\tau)
    ,\\
    \braket{0| \, \hat y_{\bm{k}}(\tau) \, \hat y^\dag_{\bm{k}}(\tau) \, |0}
    &=
    \frac{1}{2k} \left|u_{k}(\tau) + v_{k}(\tau)\right|^2
    ,
\end{align}
\end{subequations}
where, for simplicity of presentation, we have employed box-regularization in these expressions to suppress the Dirac delta distribution.
Dynamic scale factors are required to generate nonzero $v_{k}(\tau)$ in \cref{eq:Bogo_EOM}. Hence from \cref{eq:2pt} we see that inflationary expansion drives both particle production and scalar field fluctuations beyond the zero-point value.

\textit{Experimental setup.} We have seen that the dynamics of the cosmology model are encoded in the dynamics of parametric oscillators.
We now demonstrate that synthetic mechanical lattices \cite{Brandenbourger2019,Anandwade2023,Singhal2023,Tian2024} are ideally suited to solve such systems of equations.
The experimental system consists of a pair of mechanical oscillators. Each oscillator, a mass held between two springs, is equipped with an accelerometer that enables real-time measurements. The oscillator mass also features a dipole magnet that sits at the center of an anti-Helmholtz coil pair.
By driving coil currents that depend on the real-time measurements, we enact measurement-based feedback forces~\cite{Supplemental_Material} as depicted in \cref{fig:mechanical_cosmology}.

Without additional forces, the equations of motion for the position and momentum of the $n$-th oscillator at time $t$ in the experiment are $ \dot{x}_n(t) =  p_n(t) / m $ and $ \dot{p}_n(t) = - m \omega^2 x_n(t)$, where $m$ and $\omega$ are the mass and angular frequency of each oscillator with overdots representing derivatives with respect to time.
The oscillators are synthetically coupled to one another using measurement-based feedback forces.
Monitoring the real-time acceleration of each oscillator and numerically differentiating this data to obtain its jerk enables us to engineer custom forces that take into account the current state of the system \cite{Anandwade2023,Singhal2023,Tian2024}.
For this reason, it is convenient to instead work with the acceleration and jerk of each oscillator, which we will denote by $X_n(t)$ and $P_n(t)$.
With the inclusion of measurement-based feedback forces, their equations of motion are given by:
\begin{subequations}
\begin{align}
    \dot{X}_n(t) &= P_n(t)
    , \\
    \dot{P}_n(t) &= - \omega^2 X_n(t) + F_n(t,\{X_n,P_n\})
    ,
\end{align}
\end{subequations}
where $F_n$ is the feedback ``force'' on the $n$-th oscillator \cite{Supplemental_Material}.
Formally changing variables to the complex function $A_n(t) \equiv \sqrt{\frac{\omega}{2}} X_n(t) + i \frac{1}{\sqrt{2\omega}} P_n(t)$,
one finds that the synthetic mechanical lattice, after being prepared in some initial configuration, evolves in time according to:
\begin{align}
    \label{eq:synthetic_lattice_EOM}
    i \dot{A}_n(t) = \omega A_n(t) - \frac{1}{\sqrt{2\omega}} F_n(t,\{A_n^*,A_n\})
    .
\end{align}
It is from these equations that we find a natural mapping between the cosmological and experimental systems.

To perform an analog simulation of the cosmology model, here we establish a mapping between the Bogoliubov equations, \cref{eq:Bogo_EOM}, and the equations which govern the synthetic mechanical lattice, \cref{eq:synthetic_lattice_EOM}.
Since each observable discussed previously in the cosmological context depends only on various combinations of the Bogoliubov coefficients, this mapping allows us to directly construct analogs of physical quantities in cosmology using the experimental output data (\cref{fig:mechanical_cosmology}).
Because each oscillator encodes a single complex variable, we only require two oscillators, $A_1$ and $A_2$, to simulate the dynamics of the Bogoliubov coefficients $u_{k}$ and $v_{k}$ for a given momentum $k$.
We interpret the physical time in the experiment $t$ as the conformal time $\tau$ in the Bogoliubov equations and identify experimental feedback ``forces"
$
    F_1/\sqrt{2\omega} = - k A_1 - i (a'/a) A_2 + \text{c.c}
$
and
$
    F_2/\sqrt{2\omega} = k A_2 - i (a'/a) A_1 + \text{c.c}
$.
A separation of timescales between the characteristic period of oscillations and the dynamics we simulate means ``counter-rotating'' terms can be neglected in the rotating-wave approximation \cite{Salerno2014,Salerno2016,Supplemental_Material}, and the Bogoliubov coefficients $u_{k}$ and $v_{k}$ can be extracted from $A_1$ and $A_2$ respectively by amplitude demodulating the signals with carrier frequency $\omega$.

\begin{figure}[htp!]
    \includegraphics[width=0.45\textwidth]{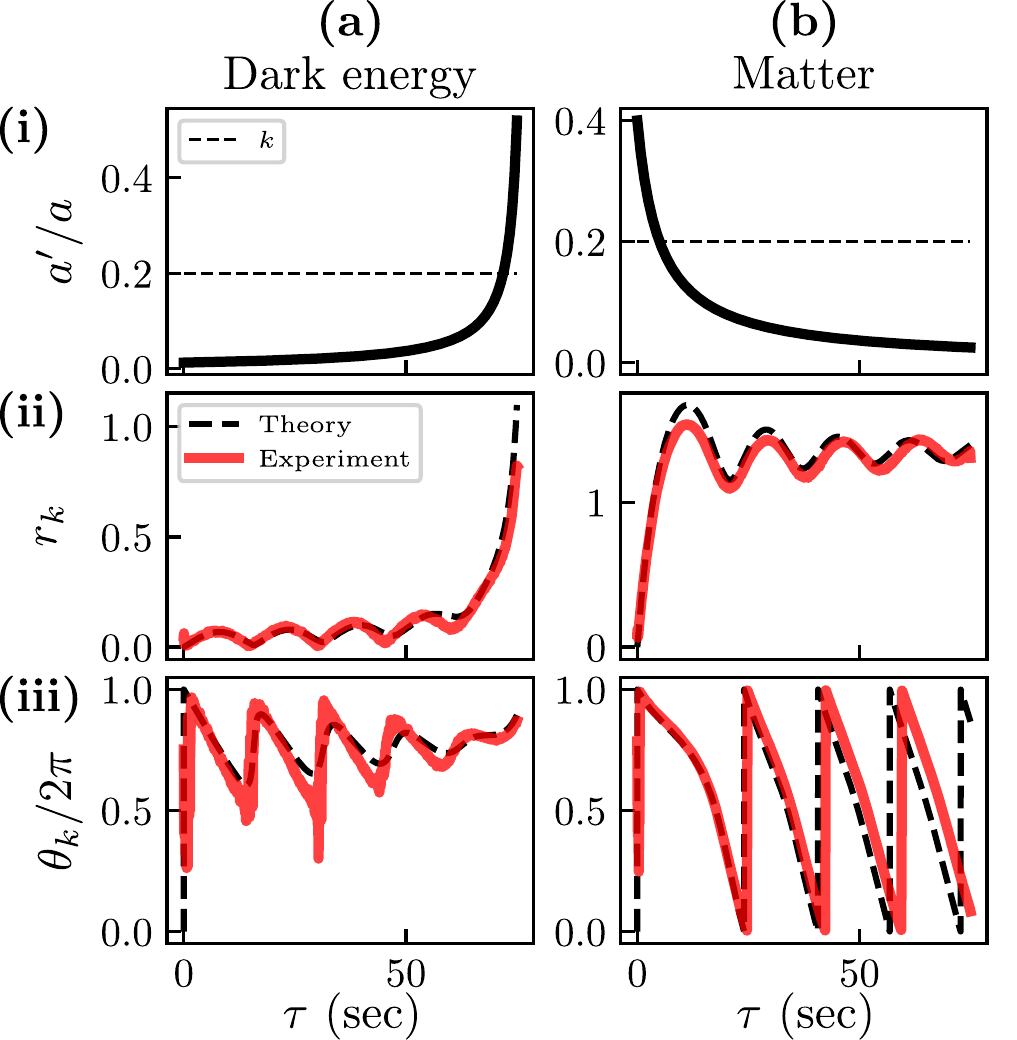}
    \caption{
    (Color online)
    Mechanical cosmology for expanding FLRW spacetimes. Columns (a) and (b) correspond to dark energy- and matter-driven expansions respectively.
    In all images, theoretical curves are shown in black and experimental results in red.
    We choose wavevector magnitude $k=0.2$ and set the scale factor according to \cref{eq:scale_factors}, with $\tilde{\tau} = 76.9 $ in  the dark energy-driven expansion and $\tilde{\tau} = 5$ in the matter-driven expansion.
    Row (i) shows the energy scales that enter into the Bogoliubov equations and appear as couplings in the measurement-based feedback forces; in particular, the off-diagonal $a'/a$ energy scale is shown in bold and the wavevector as a dashed line.
    The remaining rows show properties of the time evolved initial vacuum state with theory curves, obtained by numerically solving \cref{eq:Bogo_EOM}, as black dashed lines and experimental results, obtained by analog simulations realizing \cref{eq:synthetic_lattice_EOM}, as red solid lines.
    Row (ii) shows the squeeze parameter, expressed as $\arsinh(|v_{k}|)$, and row (iii) shows the squeeze angle, $\arg(u_{k})-\arg(v_{k})$ (mod $2\pi$), divided by $2\pi$.
    The experimental data has been amplitude demodulated with carrier frequency $13.06 \text{ Hz}$ and a $0.1 \text{ sec}$ moving average window has been applied.
    }
    \label{fig:mechanical_cosmology_exp_squeeze}
\end{figure}

\textit{Analog simulation of FLRW cosmology.} To perform an analog simulation of the cosmology model, we set a wavevector magnitude of interest, choose a functional form for the scale factor, and allow the system to evolve according to \cref{eq:synthetic_lattice_EOM} with the appropriate feedback forces.
From the output acceleration data of the experiments, we perform amplitude demodulation of the signal and construct the Bogoliubov coefficients from which various physically meaningful quantities from the cosmologic perspective can be extracted.

Here we perform analog simulations of FLRW cosmology using scale factors that correspond to dark energy-driven expansions as well as matter-driven expansions.
Setting the initial conformal time to zero for convenience gives the following conformal-time Hubble parameter in each expansion scenario:
\begin{subequations}
\label{eq:scale_factors}
\begin{alignat}{2}
    \text{Dark Energy: } & \frac{a'(\tau)}{a(\tau)} = 
    \frac{1}{\tilde{\tau} - \tau}
    &&\text{ , } 0 \le \tau <\tilde{\tau}
    ,\\
    \text{Matter: } & \frac{a'(\tau)}{a(\tau)} = 
    \frac{2}{\tilde{\tau} + \tau}
    &&\text{ , } 0 \le \tau < \infty
    ,
\end{alignat}
\end{subequations}
where $\tilde{\tau}$ is an expansion-specific parameter related to initial conditions \cite{Supplemental_Material}.
For our proof-of-principle demonstration, we choose values of $\tilde{\tau}$ appropriate to observe non-trivial dynamics on the order of $\sim 1$ minute of analog simulation.
The resulting energy scales in the Bogoliubov equations (i.e. couplings in the measurement-based feedback forces) are shown in row (i) of \cref{fig:mechanical_cosmology_exp_squeeze}.
In the matter-driven expansion, the instantaneous eigenvalues of the matrix in \cref{eq:Bogo_EOM} start out imaginary ($k<a'/a$), offering a mechanism to increase the squeeze parameter and the production of particles and fluctuations, but become real as time progresses.
The dark energy-driven expansion exhibits the opposite behavior as $a'/a$ only increases with conformal time.

\begin{figure}[htp!]
    \includegraphics[width=0.45\textwidth]{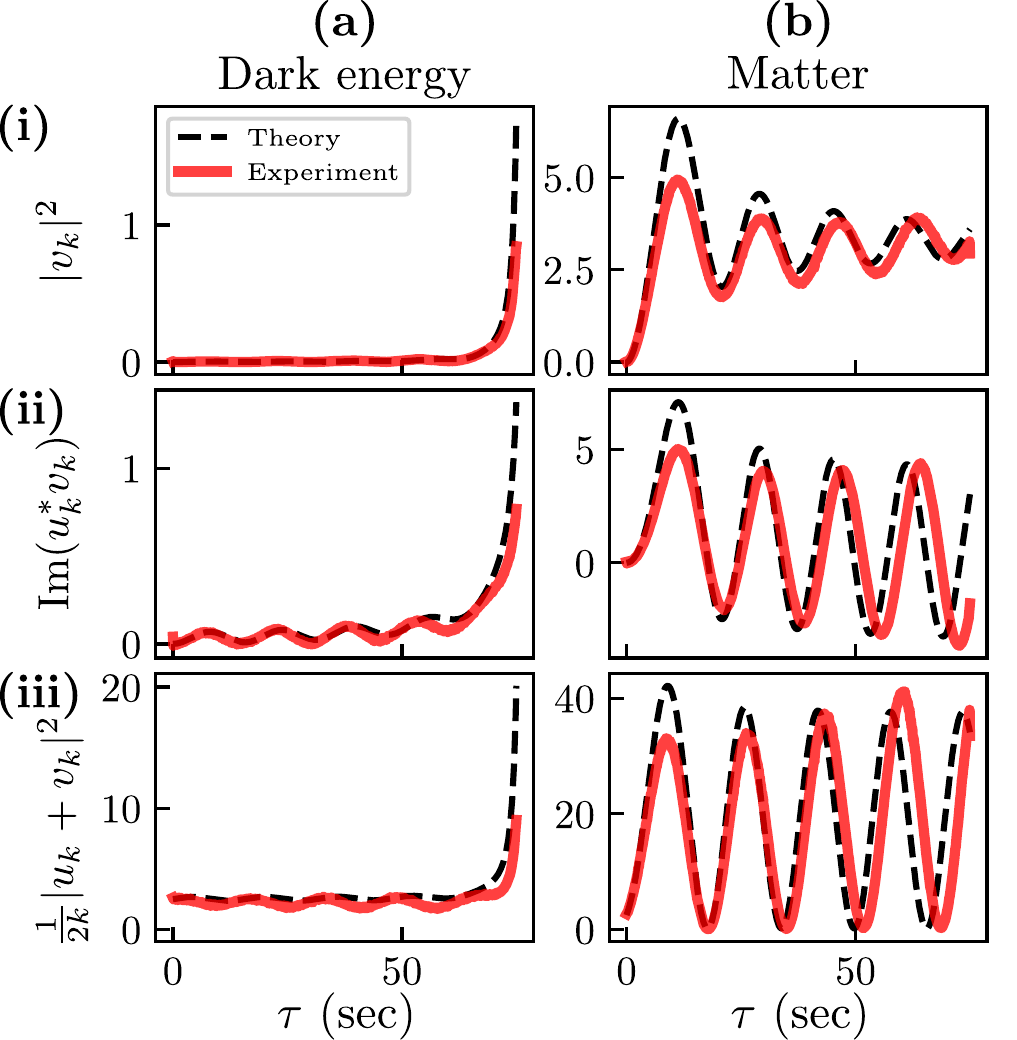}
    \caption{
    (Color online)
    Mechanical cosmology for various expanding FLRW spacetimes continued from \cref{fig:mechanical_cosmology_exp_squeeze}. Once again, columns (a) and (b) correspond to dark energy- and matter-driven expansions respectively with theory curves shown as black dashed lines and experimental results shown as red solid lines.
    The rows show the equal-time vacuum expectation values expressed in \cref{eq:2pt}:
    (i) shows the number of excitations in a $\bm{k}$ mode with magnitude $k$, (ii) shows the \textit{imaginary} part of opposite-momentum pairs produced, and (iii) shows the comoving field power spectrum.
    }
    \label{fig:mechanical_cosmology_exp_vev}
\end{figure}

The experimental results of our analog simulations of FLRW cosmology under each expansion scenario are displayed in \cref{fig:mechanical_cosmology_exp_squeeze,fig:mechanical_cosmology_exp_vev}, showing measurements for the two-mode squeezed vacuum state and vacuum expectation values, respectively.
To compare with theoretical predictions, numerical solutions to the Bogoliubov equations are also provided in each case.
As both expansion scenarios unfold, we observe the analogs of two-mode squeezing, particle production, and enhanced scalar field fluctuations.
In most cases, the experimental data qualitatively agree with the theoretical results.
Namely, the frequency of oscillations and relative heights of local minima and maxima in the experimental data are fairly consistent with theory curves across various quantities.
One of the more sensitive measures is the squeeze angle, which depends on the phase difference between the Bogoliubov coefficients.
In particular, the experimental measurement of the squeeze angle extracted from the analog simulation of the dark energy-driven expansion is somewhat noisy in relation to the theoretical curve, 
as this relative phase becomes ill-defined when the population of the second oscillator ($|A_2|^2=|v_k|^2$) gets close to zero.

Another source of disagreement comes from the normalization of the Bogoliubov coefficients, which in theory is $1 = |u_{k}(\tau)|^2 - |v_{k}(\tau)|^2$. One can use this in the FLRW models to re-express physical quantities in various ways.
This normalization is however not guaranteed in the presence of experimental imperfections.
In practice we found greater noise on the oscillator simulating $u_{k}(\tau)$.
For better agreement then, we chose to extract the squeeze parameter in \cref{fig:mechanical_cosmology_exp_squeeze} using the data from the oscillator simulating $v_{k}(\tau)$.

\textit{Conclusion and outlook.} In this Letter, we have demonstrated how synthetic mechanical lattices offer precise parallels for simulating inflationary and, more generally, FLRW cosmology.
Through mapping the bosonic dynamics of scalar field fluctuations onto the motion of mechanical oscillators, we were able to simulate and measure key physical inflationary quantities, such as particle production and the power spectrum.
This proof-of-principle study potentially opens up an entire toolbox for analog gravity in the realm of the early Universe.
Here, we have but exploited the physics of two coupled oscillators in a much more powerful system which currently contains eighteen oscillators in which one can achieve arbitrary connectivity and nonlinearity.
The scope in this experimental system is thus vast for simulating more complex actions and spacetimes,
including interaction effects such as those considered in cosmological collider physics \cite{Arkani-Hamed2015} coupling the inflaton to other heavy degrees of freedom, as well as dissipation effects and more.
Given the customizable nature of the experiment, with prospects for entering nonlinear regimes and coupling multiple degrees of freedom, there is also the potential to study dynamics beyond that of the early Universe performing analog simulations into further astrophysical domains and other branches of physics.
Synthetic mechanical lattices thus embody a highly tunable playground for simulating different cosmological scenarios to complement theoretical and observational astrophysics, as probes continue to reveal more insights and mysteries about our primordial Universe.

\textit{Acknowledgments.} We thank Naceur Gaaloul and Zain Mehdi for insightful discussions.
We gratefully acknowledge support by the National Aeronautics and Space Administration under Jet Propulsion Laboratory Research Support Agreement No. 1699891 (B.R. and S.V.),  by the United States Department of Energy, DE-SC0015655 (P.A.), and the AFOSR MURI program under agreement number FA9550-22-1-0339 (I.V. and B.G.).

\bibliography{references}

\end{document}